\documentclass{article}

\usepackage{natbib}   %% \citep{} does not work unless natbib included!
\usepackage{graphicx}
\usepackage[utf8]{inputenc}
\usepackage{rotating}
\usepackage{lineno}
\usepackage{array}

\begin{document}

\title{Non-stationarity in Esbjerg sea-level return levels: Applications in climate change adaptation\footnote{Manuscript submitted to special issue of EXTREMES, 2019}}

\author{Peter Thejll$^1$, Peter Guttorp$^3$, Martin Drews$^2$, Torben Schmith$^1$,\\ Thordis Thorarinsdottir$^3$,  Jacob Woge Nielsen$^1$,  Mads Hvid Ribergaard$^1$}

\date{%
    $^1$Danish Meteorological Institute, Lyngbyvej 100, DK-2100 Copenhagen Ø, Denmark\\%
    $^2$Technical University of Denmark, Produktionstorvet, building 424, DK-2800 Kgs. Lyngby, Denmark\\%
    $^3$Norsk Regnesentral, P.O. Box 114, Blindern, NO-0314 Oslo, Norway\\[2ex]%
    \today
}

\maketitle
%\linenumbers

\begin{abstract}

Non-stationary time series modelling is applied to long tidal records from Esbjerg, Denmark, and coupled to climate change projections for sea-level and storminess, to produce projections of likely future sea-level maxima.

The model has several components: nonstationary models for mean sea-level,  tides and
 extremes of residuals of sea level above tide level. The extreme value model (at least on an
annual scale) has location-parameter dependent on mean sea level. Using the methodology of~\cite{Bolin2015} and~\cite{Guttorp2014} we calculate, using CMIP5 climate models, projections for mean sea level with
attendant uncertainty. We simulate annual maxima in two ways - one method uses the empirically fitted non-stationary generalized extreme-value-distribution (GEV) of 20th century annual maxima  projected forward based on msl-projections, and the other has a stationary approach to extremes. We then consider return levels and the increase in these from AD 2000 to AD 2100. 

We find that the median of annual maxima with return period 100 years, taking into account all the nonstationarities, in year 2100 is 6.5 m above current mean sea level (start of 21$^{st}$ C) levels.

%\keywords{Tide gauges \and statistics \and storm surges \and climate change \and non-stationarity}
%\PACS{91.10.Tq}
%\subclass{62G32}

\end{abstract}

\section{Introduction}~\label{Intro}

As global heating proceeds into the 21st century, sea levels will rise. This rise is causing increased concerns along coast-lines globally. Not only is the rise of mean sea level ($msl$ from now on) a problem, but storm surges and even tidal range can depend on the amount of mean rise. Denmark is a flat country with a long coastline, and for mitigation and adaptation purposes estimates of the occurrence rates of high storm surges are required. We describe here a method for estimating such changes in storm surges, given a unique data set of hourly tide gauge recordings from the North Sea harbour town Esbjerg.

The adaptation approaches from local governments can be of different kinds. For example, the Esbjerg harbor has declared itself "Future-proof'', in that its facilities
are placed 4.6 metres above present mean sea level \citep{esbjerg2016}. This was considered safe based on the IPCC 4 prediction for scenario A2 of a mean sea level rise of 15-70 cm by 2100. Using the IPCC5 model data, we shall show that the 5-95\%-ile spread of our projections for Esbjerg $msl$ under scenario 8.5, which emissions so far seem to be close to, is 59-97 cm (5-95\%-ile). On top of that we must consider increasing tidal range, and extreme storm surges.

The analysis presented in this paper will be based on a statistical downscaling method for referencing climate scenario projections to local conditions. Our method allows for an ensemble approach which will allow full analysis of uncertainties in projections of future sea-levels and sea-level extremes. Briefly, we use a tested method for adjusting projections for mean sea level to observed conditions locally through regression, and then add to these projections a non-stationary tidal model and distributions of extremes drawn from an empirical extreme-value distribution.

Previous research has employed similar tools. For example, \citet{Tebaldi2012} used sea level projections due to \citet{Rahmstorf2011}, and added extreme values fitted to the excess over highest high tide, while \citet{Strauss2012} used a tidal model together with a detailed elevation map to estimate inundations relative to mean high tide. Neither approach takes into account all the nonstationarities that we consider, and their estimates therefore are likely to be somewhat conservative.
\subsection{Physics of tidal evolution}
The North Sea is a semi-enclosed marginal sea, connected to the North Atlantic at its northern flank and through the relatively narrow English Channel. Water depths are largest in the northern parts and gradually decrease to moderate values of 15-30 m in the southern parts.  Currents are generally clockwise in the region and dominated by the semi-diurnal tide system. The amplitude of the tides and the currents are strongest near the coast and vanishes in the middle of the North Sea. The North Sea is generally influenced by westerly winds. Severe wind conditions with associated storm surges are predominantly caused by low pressure systems that pass eastward over southern or central Scandinavia which occur in the winter half-year.

\begin{figure}[htp]
    \centering
     \includegraphics[width=0.98\linewidth,clip]{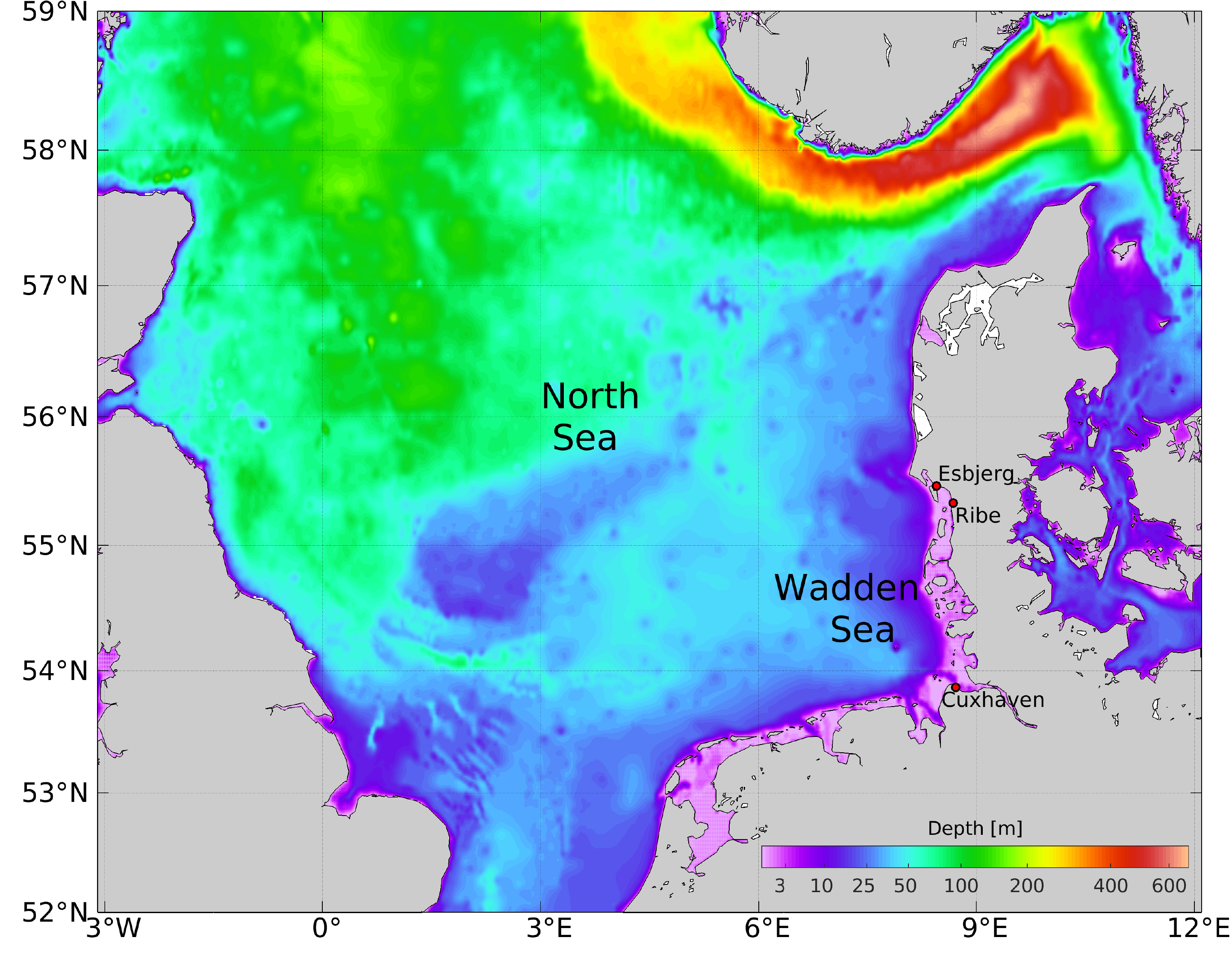}
    \caption{
    Map of Denmark and adjoining seas, with bathymetry. ETOPO1 1-arcminute bathymetry used~\citep{ETOPO1}.}
    \label{fig:mapofDenmark}
\end{figure}

The shore and ocean bottom near Esbjerg--the regularly dredged harbour sailing channel, and the shallow Wadden Sea (see Figure~\ref{fig:mapofDenmark}) outside the approach--are gently sloping, which produces a tide with a pronounced semi-diurnal tidal range of some 1.5 meters. 
The astronomical forcing and the water depth are two factors dominating the tide in the Wadden Sea, and we can expect any changes over time in mean sea level and local bottom-depth to affect the tide.

The actual influence on the tide is a balance between the effects of friction along the bottom and shoaling, and the resulting outcome due to water-depth changes, is in practise not possible to predict from theory. Empirically, we can study what is happening to tidal characteristics and relate them to known changes in conditions, and then extrapolate into the future.  Non-stationary statistical modelling of the observed tidal record will be one corner-stone of the present work--projections into the future based on a non-stationary time-series model, and climate model-based sea-level projections will then enable a view of future sea-level conditions at the coast of western Denmark.

Dynamic oceanographic modelling would require detailed knowledge of the conditions on the bottom of the sea in the Wadden Sea in the future. These would be hard to specify in order to provide accurate modelling results.
\subsection{Climate change}

~\cite{Grinsted2015} evaluated sea-level rise projections, using CMIP5 models and RCP 8.5 emission scenarios. Their global $msl$ rise estimate for 2100 is 45-80-183 cm for the 5, 50, and 95 percentiles, respectively. They also estimated local $msl$ rise to the end of the 21st century regionally, and found for Esbjerg 38-77-171 cm, respectively. 

A similar approach is used by~\cite{Bamber2019} who estimate a global $msl$ rise of 111 cm in median, with 5 and 95 percentiles at 62 and 238 cm, respectively. If the Esbjerg-to-global differences are the same as in the Grinsted et al study we might expect Esbjerg's 5, 50 and 95 percentiles to land at 55, 108 and 226 cm. 

We do not consider the contributions to $msl$ rise, or the uncertainties thereof, due to accelerated melting of ice on land.

\section{Data}
\subsection{Tide-gauges}

In this paper we use hourly tide-gauge data observed at Esbjerg, Denmark on the North Sea coast.
At DMI, such records have been maintained for the Esbjerg station since January 1 1891. 
%For Esbjerg the relevant tide-gauge station numbers are 25148 and 25149. Farvandsvæsenet used to operate a gauge with station number 25146, but we do not rely on that data here, and Kystdirektoratet operate a gauge since 1970 with the station number 25147, which we also do not use here.
 
Hourly data for Esbjerg (on the hour) were taken from the database of tide-gauge readings held at DMI. 
The complete set of hourly Esbjerg data has not been published before, but a shorter record, starting in 1951, is available at the GESLA database site~\citep{Woodworth2016}. Relevant technical notes regarding the data held at DMI are available in~\cite{Hansen2018}.

\begin{figure}[htp]
    \centering
\includegraphics[width=0.96\textwidth]{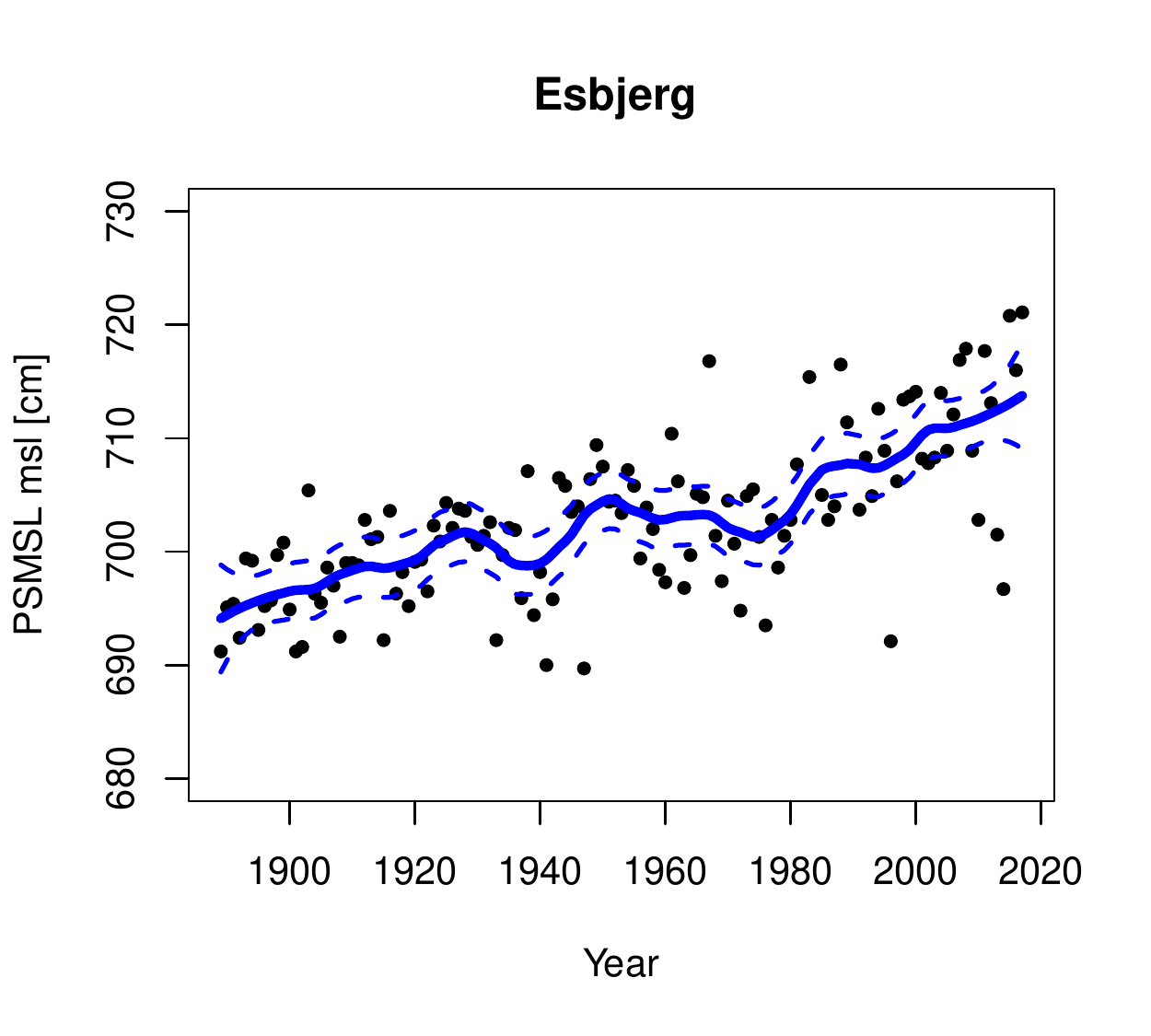}
    \caption{Annual PSMSL mean sea levels for Esbjerg.
    The smoothing curve is a \textit{loess} curve~\citep{ClevelandGrosse92-b}, with span=0.25. The dashed lines show, respectively, the 5 and 95 percentile confidence levels for the smoothed blue curve. Data from PSMSL downloaded May 4 2019.}
    \label{fig:msl_plot_Esbjerg}
\end{figure}
\subsection{Quality Issues}

Our analysis will require accurate mean levels and reliable extremes. For mean sea level we use the Permanent Service for Mean Sea Level (PSMSL) data~\citep{PSMSL2012}--see Figure~\ref{fig:msl_plot_Esbjerg}. DMI provides the quality-checked data (see~\citet{Hansen2018}) used by the PSMSL for Danish stations, but the hourly records themselves will not here be used for deriving mean levels--just for the extremes. 
 
The reference levels of the data are as given by the PSMSL for the annual data, while a local reference is used for the hourly data. Our analysis method for extremes removes the vertical reference and first-order trends in the data.

It became evident from the hourly data for Esbjerg, that a problem with the chart-recorder used in the early days prevented recording extreme high sea-level. Inspection of the distribution of the extreme highs had revealed that until 1910, realistically high extremes were absent. We shall therefore only perform analysis on data from 1910 and to the end of 2018.

\subsection{Climate data for $msl$ evolution}
Mean sea level is modelled using temperature data and projections from CMIP5 models~\citep{Taylor2012}.
We use the 33 climate models from CMIP5 that calculate both historical estimates and projections for reference concentration pathway (RCP) 8.5 used in the experiment. 

\section{Methods}
\subsection{Sea level projections}~\label{slp}

The climate projections in CMIP5 generally do not report sea level rises directly. 
The IPCC AR5 \citep[ch. 13]{ar5} sea level modelling uses steric increases from CMIP5 projections, and sub-gridscale land ice models driven by CMIP5 temperature projections to calculate sea level rises.
Our projection of local sea level is an alternative method based on the statistical downscaling technique by \cite{Guttorp2014} and \cite{Bolin2015}. 
On an annual time scale, historical global sea level is related by time series regression to the corresponding global temperature. In turn the local sea level, adjusted for glacial rebound, is then related to global sea level, also by time series regression. 
In order to obtain projections, the global sea level uses the regression model with projected global temperatures replacing the observed temperatures. The local sea level is then obtained by computing regression estimates of glacially adjusted sea level, and adding the inverse adjustment assuming a constant annual glacial rebound rate. 

In order to compute simultaneous confidence bands for the sea level rise projections we take into account the ensemble spread of the climate models, the uncertainty of the two regressions, and the uncertainty of the glacial isostatic adjustment. The \textbf{R} package \textit{excursions}~\citep{Bolinb,Bolinc,Bolina} has the functions needed to compute the bands.

\subsection{Tides analysis}
Time series of sea level observations are typically analysed by harmonic analysis. The result is amplitudes of constituents with different periods. Having deconstructed an observed series we can separate the harmonic part which is astronomical in origin from the stochastic part which is due to factor such as storm surges, and waves.

The \textit{ftide()} function from the \textit{TideHarmonics} library~\citep{Alec2017}, written in R~\citep{R2018}, is used to fit, and then remove, as much as possible of the tidal signals present in the tide-gauge series.
 \textit{ftide()} includes user-selectable sets of tidal harmonics,  and allows for removal of the lunar nodal oscillation (period 18.6 years) and a smooth background term with periods longer than the annual periodicity. The sets of harmonics are based on the PSMSL TASK-2000 software~\citep{task2000}. The smoothing used for the background term is based on the \textit{loess} procedure~\citep{ClevelandGrosse92-b}, and uses a user-selected parameter for the degree of smoothing. We selected $\alpha=0.25$, after testing for robustness in the results.

We will fit one year of data at the time. It is also possible to fit all the years of hourly data (near one million values), but we found that this does not allow the harmonic model to adapt to the data fully. Fitting all data at once implies the same amplitude and phase in each constituent along the entire series, and this is not realistic since we expect the wave dynamics to be dependent on water depth.  By employing annual fits to the data we allow the harmonic model to adapt, to a degree, to the rising sea level.

\subsection{How to determine the best harmonic constituent set}
\begin{figure}
    \centering
    \includegraphics[scale=0.69]{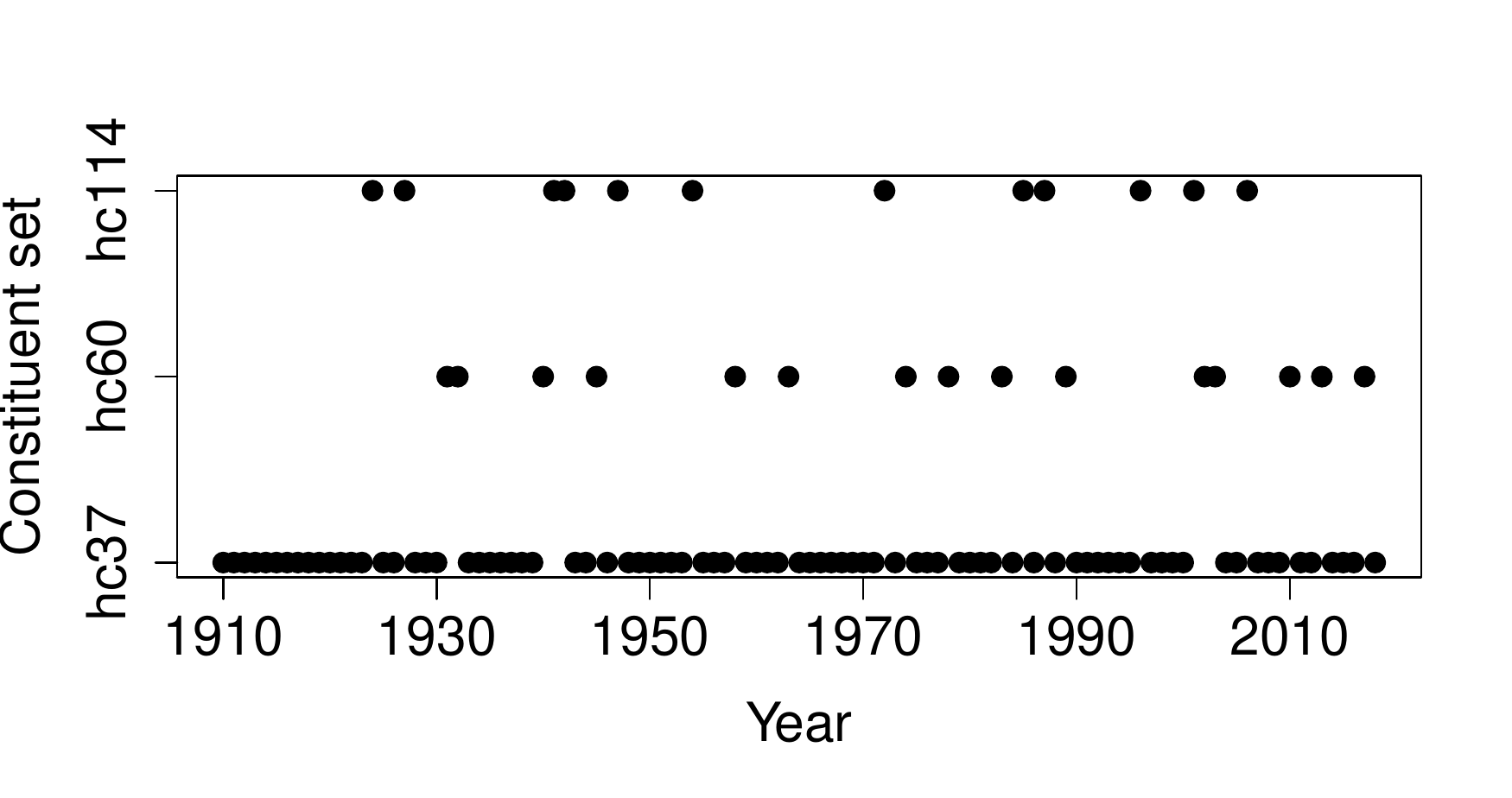}
    \caption{Which tidal constituent set is best for each year? Plotted with a circle is the constituent set that was best each year. The set "hc7" was also tested but is never the best. The distribution between the three sets appears to be uniform through the dataset.}
    \label{fig:constiset}
\end{figure}

\begin{figure}[tbp]
    \centering
    \includegraphics[width=0.9\textwidth]{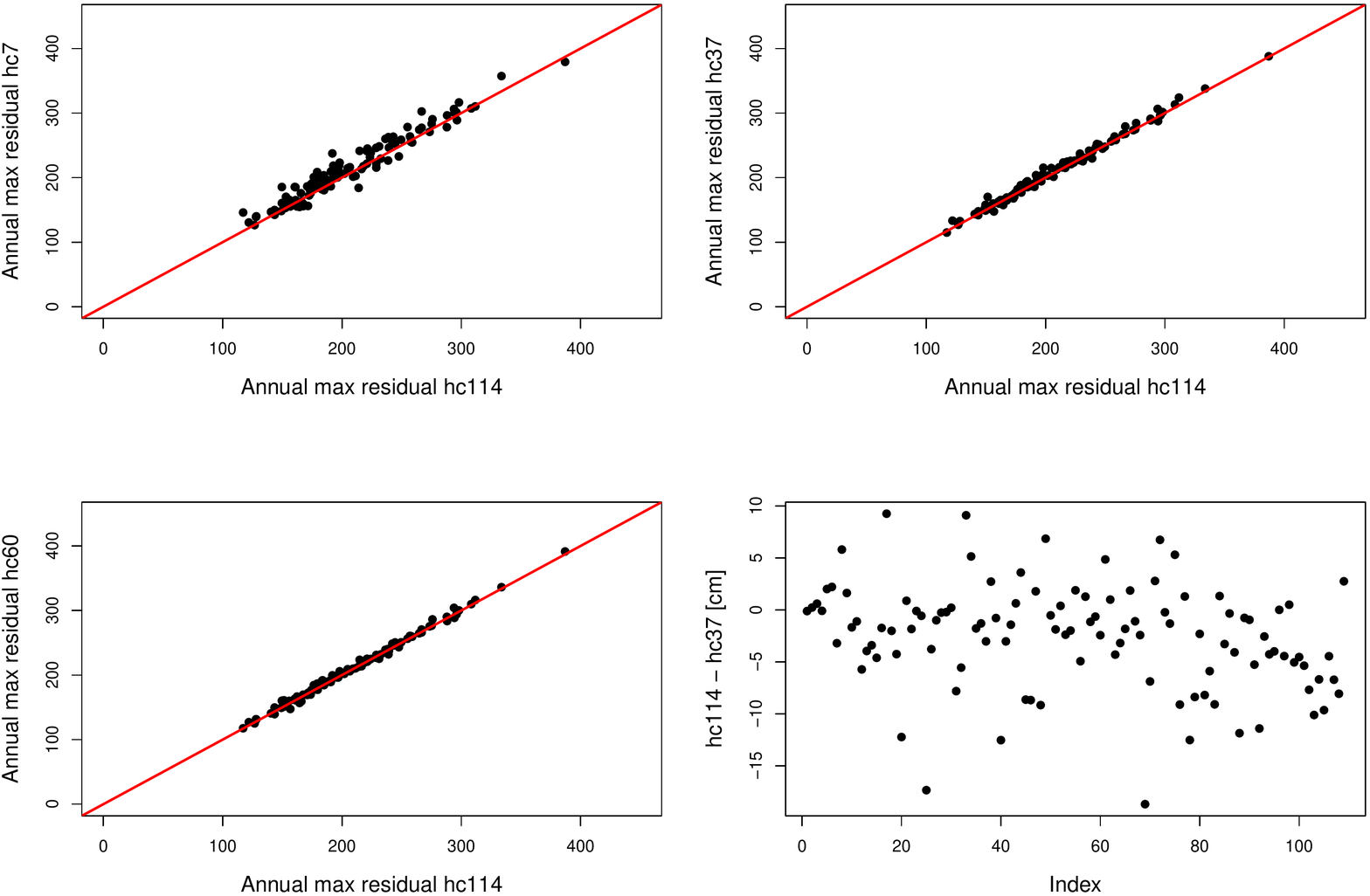}
    \caption{The annual maxima detected depend on the tidal harmonic model chosen. (\textbf{upper left:}) Annual residual maximum from the hc7 model plotted against the same from the hc114 model. (\textbf{upper right:}) hc37 against hc114, (\textbf{lower left:}) hc60 against hc114, (\textbf{lower right:}) difference between hc37 and hc114 - the standard deviation of the difference is 5 cm.}
    \label{fig:annmaxIdentified}
\end{figure}

In preliminary testing, we investigated which harmonic model to fit to the Esbjerg hourly data.  \textit{ftide()} offers many models, each consisting of different numbers, and selections, of constituents. When does over-fitting occur - when could more be gained in terms of explained variance by including more terms? We based model-choice on identifying the 'best' harmonic model, based on the Akaike Information Criterion (AIC)~\citep{Akaike1974}, across all years fitted. Table~\ref{tab:bestHCN} shows that the most frequently occurring best model, according to AIC, is the hc37 model, consisting of 37 periodic constituents with periods from less than a day up to a year, which we have used throughout this analysis. We also list the standard deviation of the fit-residuals for the models considered for an example year (2018)--the smallest residuals are found by the hc114 set of constituents, but given the AIC information the most parsimonious model set is hc37. Figure~\ref{fig:annmaxIdentified} shows that the annual residual maxima detected depend somewhat on which tidal set of constituents are fitted, and that the \textit{differences} in maxima detected have a standard deviation of 5 cm. Is the choice of hc37 for all years a good choice? For instance, does hc37 predominate only for some part of the 20th century and other constituent sets replace it as best choice? Figure~\ref{fig:constiset} shows that this is not likely to be a large problem.

\begin{table}[!htp]
    \caption{Best harmonic model (set of constituents) to choose for the Esbjerg hourly time-series, in \% of AIC-tests on annual data-segments (column 2). Testing was based on AIC. On the annual data, the smallest AIC was significantly smaller (i.e., by 6 (which is the AIC-difference corresponding to p=0.05), or more) than the second-best model in 75\% of the years, using hc37. Column 3 shows the standard deviation of the fit residuals in a single year segment (2018). The model names refer to the constituent set names defined in the \textit{TideHarmonics} \textbf{R} library.}
    \label{tab:bestHCN}
    \centering
\begin{tabular}{rcl}
\hline
Model & freq. on annual data (\%) &  $\sigma_{2018}$\\
\hline\hline
hc7 &0&36.7 \\
hc37 &75&34.6 \\
hc60 &14&34.4 \\
hc114 &11&34.2 \\
\hline
    \end{tabular}
\end{table}

\subsection{EV-analysis with non-stationarity}~\label{jgjgv}

Once the tidal signal has been removed what remains should mainly be the sea-level variations driven e.g., by the wind--that is, storm surges. These residuals could then tell us what the variability due to storms is, and could be used in forward modelling of the sea-level in a stochastic sense. Therefore, we first need to know whether the distribution of residuals appears to be stationary in the historic record, and whether external factors may be influencing the level of the surges.

Changes in storminess immediately come to mind as a potentially important cause of non-stationarity, as do changes in water depth--e.g. due to silting, dredging or, importantly, ocean sea-level rise.

We will inspect the distribution of residuals and test whether stationarity is in place, and if not, try to attribute non-stationarity to external factors.
Alternatively, we can simply take the hourly residuals as observed and re-sample them in building scenarios.

These two approaches play different roles in this paper - the first approach, using non-stationary GEV-distributions, includes and evaluates the consequences of taking non-stationarity also in the extremes into account, while the simpler approach of just sampling the empirical (and therefore stationary) residuals allows an estimation of the magnitude of the former assumption. The two approaches have consequences in several areas.

Firstly, if the residuals show strong signs of non-stationarity it becomes important for creating credible future scenarios to capture and understand the non-stationarity. Secondly, as extreme value distributions are necessarily based on small amounts of data (we only have 106 years of data from which to determine annual maximum residuals, for instance), it may give precision advantages to simply sample the historical residual distribution. In both approaches subsequent analysis of scenario properties, such as return levels or GEV-distribution parameters can be done on the projections.

\subsubsection{Testing for non-stationarity}
We will first test for stationarity in tidal residuals.
The theory of generalized extreme value distributions~\citep{Coles2001} includes the possibility to test for non-stationarity in GEV-model parameters fitted to detected extremes, and the \textbf{R} package \textit{fevd()}~\citep{Gilleland2016} will be used to perform such testing. A large amount of preliminary testing has been performed and will not be detailed here --- we will only describe testing outcomes of allowing mean sea level 
to be co-variates of the location. 
We shall again base model-selection on AIC.

\subsubsection{Establishing return levels}
Specifying return levels for given extreme events is an important way to describe risks in storm surge work. We will estimate return levels from model projections using parametric formal methods (Equation 3.4 in~\cite{Coles2001}), as implemented by the\textit{ fevd()} package. The return levels will be relative to year 2000 since that is our vertical zero in the simulations of future sea levels.

\subsection{Synthesis}

Two methods for using distributions of residuals or extremes are to be considered as explaned in Section~\ref{jgjgv}.

\subsubsection{Method 1; sea-level modelling with focus on GEV-models for annual extremes}~\label{sec:method1}
Guided by the results on which tidal constituents are stationary with respect to $msl$ and which are not, we  propose the following model for future extreme sea levels at Esbjerg:

\begin{eqnarray}
l_{ext}(t) &=& msl_t \nonumber \\
&+& \sum_{i=1}^{N_s} a_i\times\cos(\phi_t)\nonumber\\
&+& \sum_{j=1}^{N_{ns}} A_j(msl_t)\times\cos(\phi_t) \nonumber \\
&+& GEV(msl_t) \nonumber 
\end{eqnarray}
where $t$ is the year of the 'future' ($2000, \ldots, 2100$), $N_s$ is the number of stationary constituents and $i$ loops over these, $a_i$ is the amplitude of the $i$th stationary constituent, $\phi_t$ is a random phase at time $t$; $N_{ns}$ is the number of non-stationary constituents, $A_j(msl_t)$ is the  coefficient of the $j$th non-stationary constituent and depends on $msl_t$. The model for $msl_t$ is of the form $A_j = \alpha_j + \beta_j*msl_t$, with the $\alpha_j$ and $\beta_j$ empirically determined.

However, the above model ignores any fixed phase-relationships that may be in place between various tidal constituents.

Another approach, which maintains the empirical interrelationships of the phases to a greater degree, is:
\begin{eqnarray}
l_{ext}(t) &=& msl_t\nonumber\\
&+& \sum_{i=1}^{N_s} \left[a_i\times\cos(\phi_t) + b_i\times\sin(\phi_t)\right]~\label{eq:eq6}\\
&+& \sum_{j=1}^{N_{ns}} A_j(msl_t)\times\cos(\phi_t)\nonumber\\
&+& GEV(msl_t).\nonumber
\end{eqnarray}

Since the non-stationary components have evolving amplitudes, the phases between these and the stationary constituents must necessarily drift, and we do not attempt to model that relationship. In principle, the phases $\phi_t$ used for the non-stationary constituents could be different from the $\phi_t$ used for the stationary ones as long as the $\cos$ and $\sin$ in the non-stationary part of the model (Equation~\ref{eq:eq6}) received consistent values. Such elaborate schemes will not be entered into here - we will use the same $\phi_t$ random value to represent all phases for year $i$. As we will do this for all ensemble members we will effectively be sampling all the possible phases.

In details, we proceeded as follows: For AD 2000 to AD 2100, and each of the 10.000 $msl$ simulations, we constructed a tide(t) and then drew a GEV value from the non-stationary GEV-fit GEV(msl(t)), and added these up. This results in 10.000 simulations of annual maxima for each of 101 years. For the first year of these we then took 99 101-value segments and fitted a GEV, deriving parameter values and from these return levels for a set of return periods. We thus made 99 draws of internally consistent sets of return levels. Each return level for each return period thus had 99 samples. We found lower, median and upper percentiles of these and could then draw typical return level curves. We repeated this for each year. The difference between the 50 percentile values gave us expected changes in the median return level. Error propagation based on  upper and lower percentiles then gave us upper and lower uncertainty limits on the return level and we could plot the change in return levels for the set of return periods and also the upper and lower percentiles.

\subsubsection{Method 2; Sampling the residuals}~\label{sec:method2}

As explained in Section~\ref{jgjgv} we can also just sample hourly values of the 106 years of hourly residuals and add them to our msl-model and the model for the tide. This approach assumes that future sea-level residuals are distributed like the 20th century values. This is a suitable null experiment, good for highlighting the effect of assuming non-stationarity in the extremes distribution of Method 1 above, and will be performed here. The method has the advantage that it directly models future hourly  values as if they were observed--we can directly apply GEV-methods and estimate return levels. The Danish Coastal Authority (KDI from now on) estimates return levels from detrended annual sea-level maxima~\citep{kdi2017} and we can do the same with our samples.

In detail, we proceeded as follows: For each year and each of the 10.000 $msl$ simulations we constructed a tide(t) and then drew a sample from the empirical distribution of 20th century hourly residuals and added them to the msl(t) and tide(t). This resulted in 10.000 simulations of hourly sea-levels. Using bootstrap with replacement we drew 24x366 values and pretended this was a year of tide-gauge observations. We determined the maximum value and set it aside. We repeated this 106 times, thus simulating 106 annual maxima as in the observations. We then fitted a GEV-distribution to each simulated year and from the parameters determined return levels. The bootstrap with replacement was then repeated a large number of times and 5, 50, and 95 percentile return levels determined  for each return period. This was done for each year, and the change in return levels determined as in the GEV-sampling method.

\section{Results}
\subsection{Simulated rise in mean sea level}
From the 10.000 simulations of 21st C sea-level rise, calculated as described in Section~\ref{slp}, we estimate a difference in mean sea level for the last and the first decade of the 100 years simulated of 59-78-97 cm (5-50-95\%-iles).

\subsection{Trend in tides }

\begin{figure}[htp]
    \centering
\includegraphics[scale=0.7]{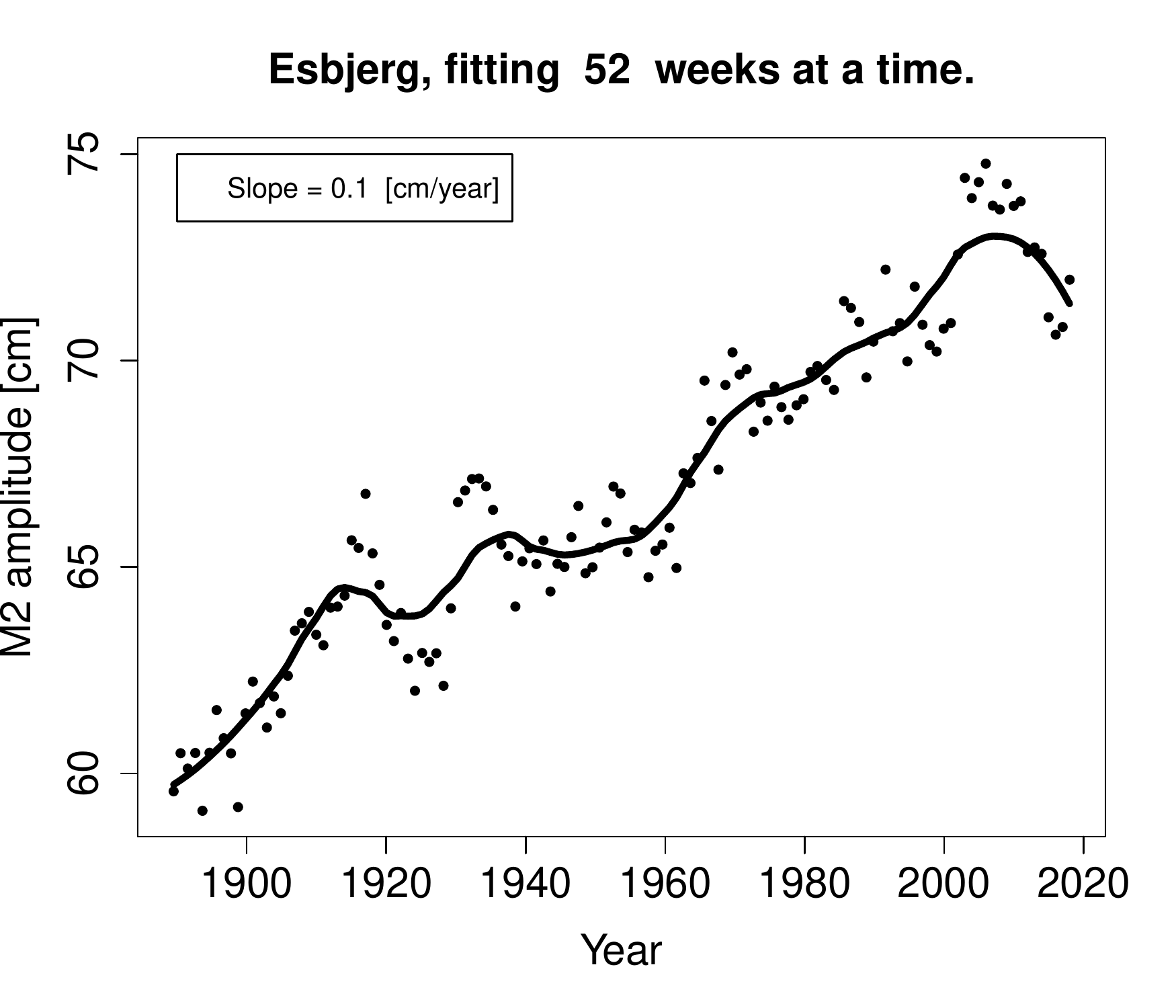}
    \caption{Illustration of the non-stationarity of the tide at Esbjerg: Amplitude [in cm] of the M2 constituent in the Esbjerg record, as revealed by fitting annual segments. There is non-stationarity--the amplitude increases with time, at about 1 mm per year. Seen is also the lunar nodal influence on the amplitude. Other constituents also have trends (not shown).}
    \label{fig:M2amplitude}
    \end{figure}

From an early stage we found that the major tidal constituent - M2 - was nonstationary. This became evident when annual segments of the time-series were fitted and the M2 amplitude was calculated and inspected. See Figure~\ref{fig:M2amplitude}.

\begin{figure}[htp]
    \centering
    \includegraphics[width=0.7\textwidth]{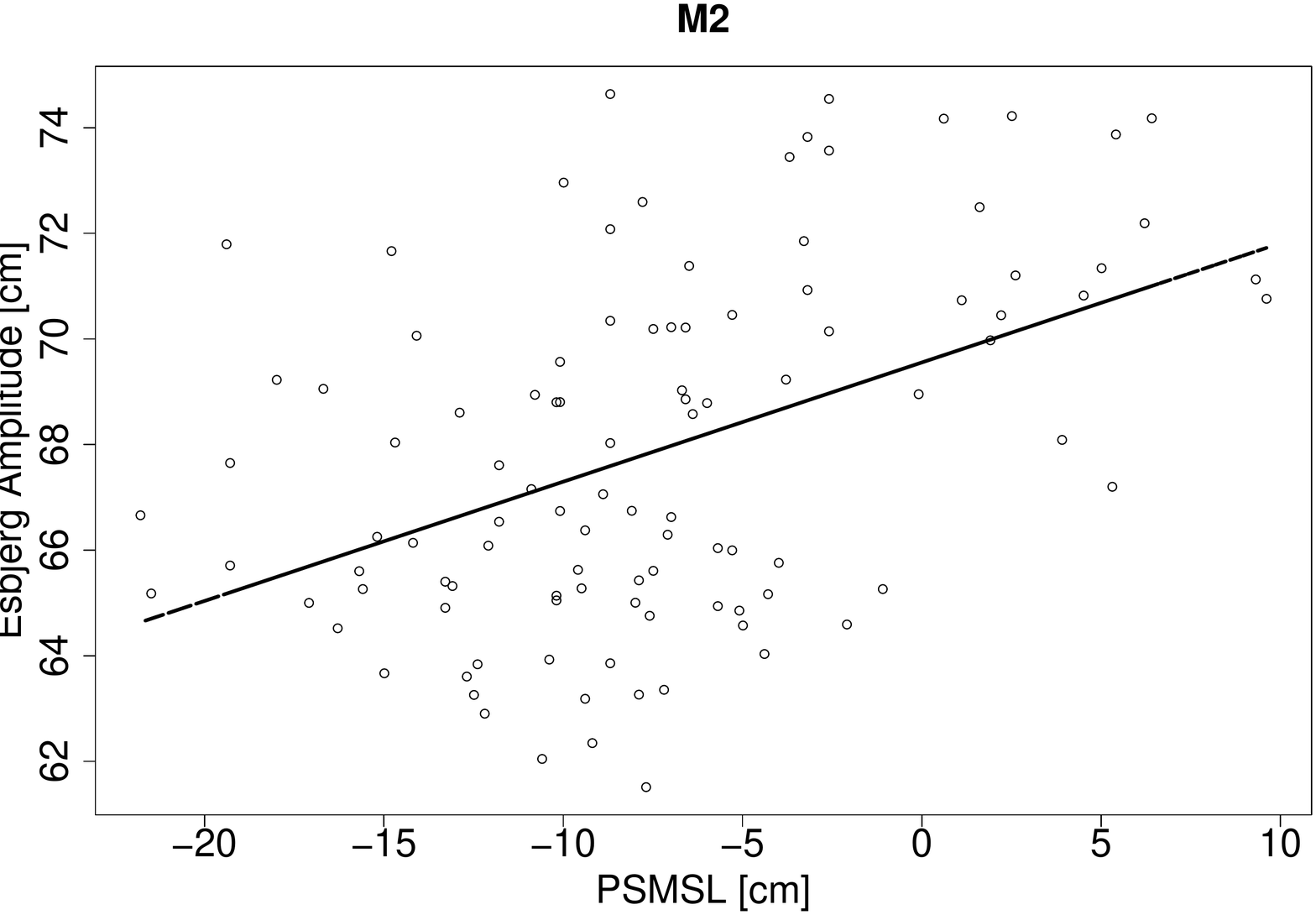}
    \caption{The annual M2-amplitude values plotted against the PSMSL mean sea level each year, for Esbjerg. The line is the least-squares linear fit. All errors were assumed to be on the ordinate.}
    \label{fig:M2_vs_PSMSL}
\end{figure}

Given this, we inspected all 37 constituents, and identified several that were clearly not stationary over the 106 years of data. While a trend against time can easily be spotted by eye, a linear relationship with PSMSL sea-level is less statistically significant (see Figure~\ref{fig:M2_vs_PSMSL}), and in the end we settled on 4 constituents (M2, S2, N2 and K2) that  clearly have cos and sine-coefficients significantly related to PSMSL. %Tables~\ref{tab:constituent_trends} %and~\ref{tab:constituent_trends_msl} show the %trends against time and PSMSL, respectively
Table~\ref{tab:constituent_trends_msl} shows the trend against msl. A table giving annual coefficients for cos and sine, too large to include here, is available on request (See Table~\ref{tab:fulltableTides}).

\begin{table}[!tbp]
\begin{center}
% Note to self: table generated with code: # "Modelling_M2_as_function_ofmsl_Esbjerg_3"
\caption{Trends of constituents against PSMSL mean sea level. The 4 top constituents have a regression slope significant at 4$\sigma$ or better.}\label{tab:constituent_trends_msl}
\begin{tabular}{rrrr}
\hline\hline
\multicolumn{1}{c}{Name}&\multicolumn{1}{c}{Trend}&\multicolumn{1}{c}{Trend err.}&\multicolumn{1}{c}{Z}\tabularnewline
\multicolumn{1}{c}{}&\multicolumn{1}{c}{cm/cm}&\multicolumn{1}{c}{cm/cm}&\multicolumn{1}{c}{}\tabularnewline
\hline
M2&0.22585&0.04374&5.2\tabularnewline
S2&0.05806&0.01284&4.5\tabularnewline
N2&0.05309&0.01168&4.5\tabularnewline
K2&0.02882&0.00718&4.0\tabularnewline
\hline 
K1&0.01574&0.00634&2.5\tabularnewline
M4&-0.00827&0.00638&-1.3\tabularnewline
O1&0.01272&0.00728&1.7\tabularnewline
M6&0.00226&0.00233&1\tabularnewline
1M1k.3&0.00156&0.00193&0.8\tabularnewline
S4&0.00144&0.00134&1.1\tabularnewline
1M1N.4&0.00047&0.0042&0.1\tabularnewline
nu2&-0.00228&0.00985&-0.2\tabularnewline
S6&-0.00244&0.00129&-1.9\tabularnewline
mu2&0.01527&0.00764&2\tabularnewline
2N2&-0.00017&0.01386&0\tabularnewline
OO1&0.00055&0.00498&0.1\tabularnewline
lam2&-0.00252&0.00968&-0.3\tabularnewline
S1&0.00357&0.00579&0.6\tabularnewline
M1&-0.00424&0.00522&-0.8\tabularnewline
J1&0.00713&0.00386&1.8\tabularnewline
Mm&0.05781&0.05387&1.1\tabularnewline
Ssa&-0.00845&0.01289&-0.7\tabularnewline
Sa&0.0024&0.00786&0.3\tabularnewline
MSf&-0.02175&0.03517&-0.6\tabularnewline
Mf&-0.07246&0.04188&-1.7\tabularnewline
rho1&0.00715&0.00514&1.4\tabularnewline
Q1&0.00361&0.00795&0.5\tabularnewline
T2&-0.00532&0.00722&-0.7\tabularnewline
R2&-0.01611&0.00663&-2.4\tabularnewline
2Q1&0.00648&0.00548&1.2\tabularnewline
P1&0.0057&0.00583&1\tabularnewline
2S.1M2&0.00092&0.00261&0.4\tabularnewline
M3&0.00284&0.00155&1.8\tabularnewline
L2&0.04078&0.01867&2.2\tabularnewline
2M.1k3&0.00167&0.00078&2.1\tabularnewline

M8&-0.00116&0.00091&-1.3\tabularnewline
1M1S.4&-0.00906&0.00386&-2.3\tabularnewline
\hline
\end{tabular}\end{center}
\end{table}

\begin{table}[!htp]
%    \centering
%    \begin{tabular}{c|c}
%         &  \\
%         & 
%    \end{tabular}
\caption{Time-dependence in tidal harmonic components. The fitted coefficient of the cos- and sine-components of each of the 37 tidal constituents are given. Row 1 gives the coefficients when fitting all years at once, while the subsequent rows give, for year 1, $\ldots$ , year 106 each annual set of coefficients. The values in the table are not shown here, but are available on request from the corresponding author at \textit{pth@dmi.dk}. 
%\footnote{Note to self: look for file   \texttt{Esbjerg\_1910\_combined\_stat\_nonstat.csv} also on ftp.thejll.com}
}\label{tab:fulltableTides}
\end{table}

\subsection{Is the extreme value distibution non-stationary?}
Table~\ref{tab:GEVresults} show the results of fitting GEV-models to annual residual maxima -- both a stationary and a non-stationary model. 
The least good fit, according to AIC, is the stationary fit, while the best is the non-stationary model with $msl$ as co-variate in location.
%The significance of the two co-variates is low, however--only %slightly above the 2$\sigma$-level for $\mu_1(msl)$ and 2 for %$\sigma(NAO)$. Allowing only non-stationarity in the form of %$\mu_1(msl)$ the significance rises slightly, and this next-best fit %(line 2) is certainly not significantly worse than the best fit %(line 4). 
%For simplicity we shall adopt fit 2 as the extreme value distribution-candidate here, ignoring the potentially interesting suggested dependence in $\sigma$ on NAO. 
The shape parameter $\xi$ never gains statistical significance, and we shall set it to zero throughout:  
\begin{eqnarray}
GEV(\mu = 195 + 1.4\cdot msl(t); \sigma=41;  \xi = 0),\label{eq:GEV1}
\end{eqnarray}
for $msl(t)$ in $cm$.

\begin{table}[!htp]
\centering
\caption{Results of fitting stationary and non-stationary GEV models to hourly Esbjerg annual-maximum residuals.}\label{tab:GEVresults}
\begin{tabular}{ccccccp{3.51cm}}\hline
No.& $\mu_0$ & $\mu_1(msl)$ & $\sigma_0$ & $\xi$ & AIC & Notes\\\hline\hline
1   & 185$\pm$5 &--             & 42$\pm$3 & -.03$\pm$.08  & 1120.4    &  Stationary GEV\\
2   & 195$\pm$6 & 1.4$\pm$.6    & 41$\pm$3 & -.01$\pm$.08  & 1116.9    & GEV non-stationary in $msl$\\

\hline
\end{tabular}
\end{table}

%\begin{table}[!htp]
%\centering
%\caption{Results of fitting stationary and non-stationary GEV models to hourly Esbjerg annual-maximum residuals.}\label{tab:GEVresults}
%\begin{tabular}{cccccccp{2.21cm}}\hline
%No.& $\mu_0$ & $\mu_1(msl)$ & $\sigma_0$ & $\sigma_1(\tiny{NAO})$& $\xi$ & AIC & Notes\\\hline\hline
%1   & 185$\pm$5 &--& 42$\pm$3    &--& -.03$\pm$.08 & 1120.4 &  Stationary GEV\\
%2   & 195$\pm$6 & 1.4$\pm$.6 & 41$\pm$3&--& -.01$\pm$.08 & 1116.9 & GEV non-stationary in $msl$\\
%3   & 186$\pm$4 &--& 41$\pm$3 & -7$\pm$4 & .007$\pm$.08 & 1119.0 & GEV non-stationary in $NAO$\\
%4   & 194$\pm$6 & 1.3$\pm$.5 & 39$\pm$3 & -6$\pm$3 & .03$\pm$.08 & 1116.2 & GEV mon-stationary in $msl$ and $NAO$\\
%\hline
%\end{tabular}
%\end{table}

\subsection{What the composite model for the future shows}
In order to fully understand the effect of including non-stationarity in the GEV-distribution of extremes we now show results from the two approaches to residual representation (Sections~\ref{sec:method1} and~\ref{sec:method2}). 

\subsubsection{Return level changes based on modelling of non-stationarity in GEV models}
\begin{figure}[!htp]
    \centering
\includegraphics[width=0.48\textwidth]{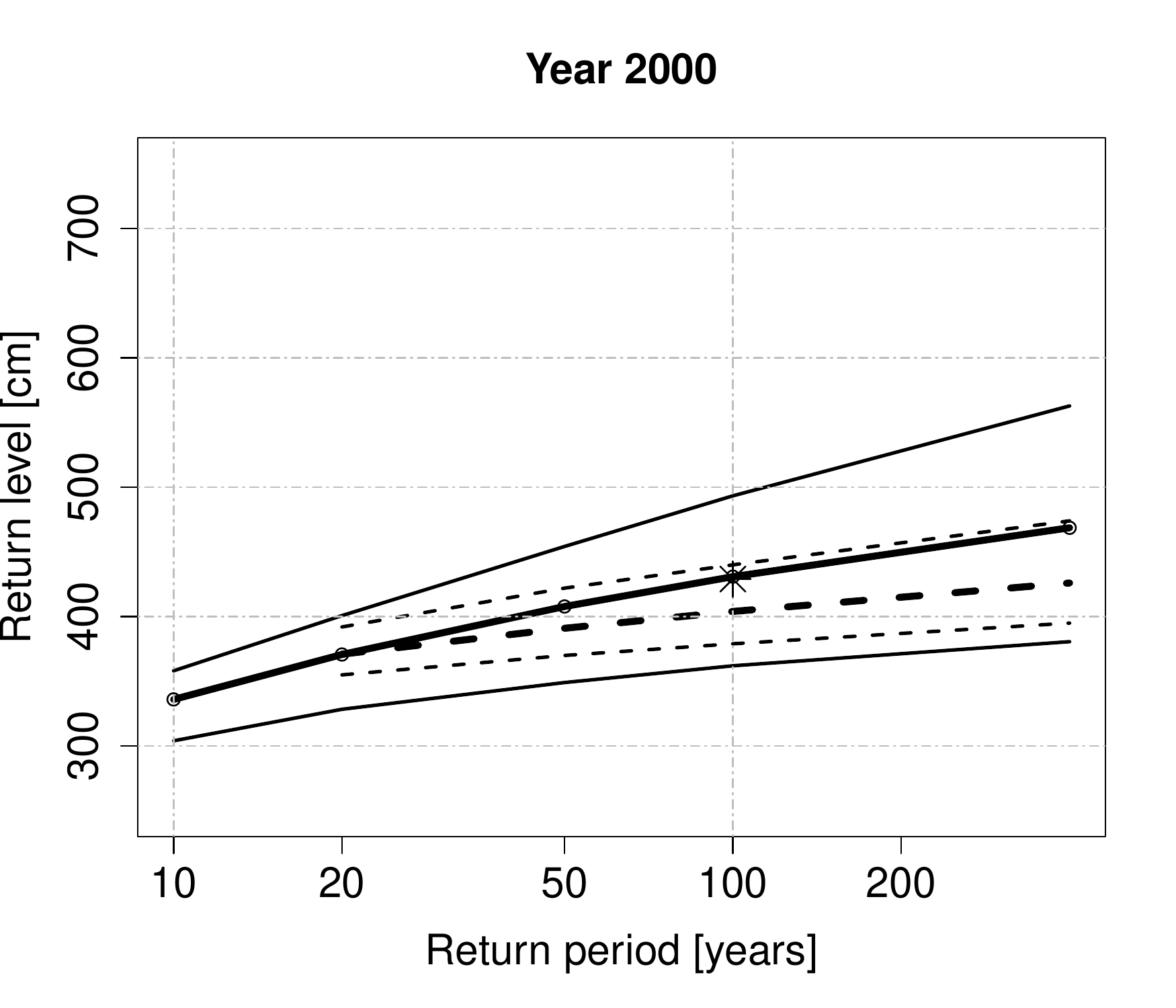}
\includegraphics[width=0.48\textwidth]{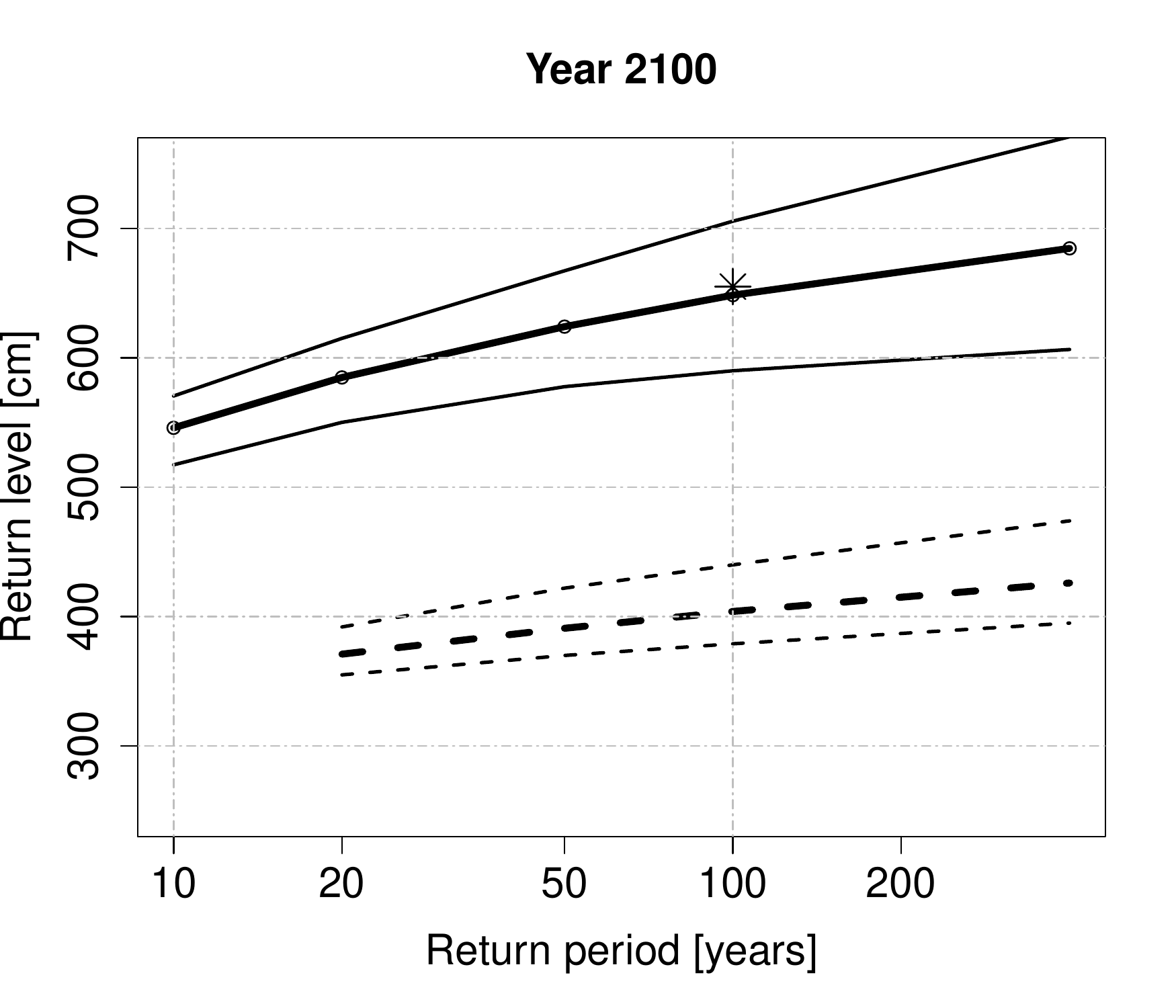}
\includegraphics[width=0.69\textwidth]{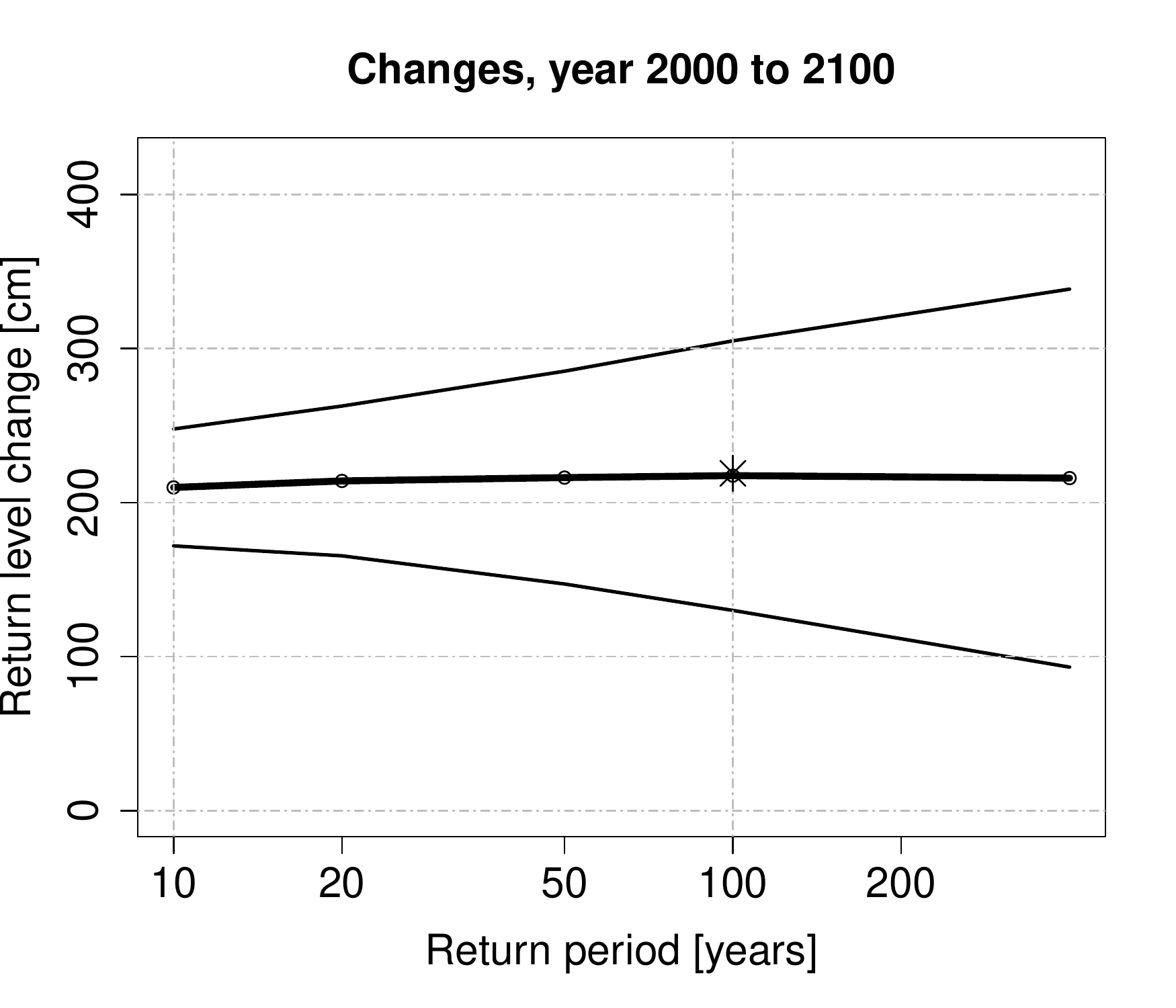}
    \caption{(Upper panels:) Return levels for years AD 2000 and AD 2100 of the scenario period. Solid lines show return levels calculated from simulated annual maxima and GEV-representations. Dashed lines show the KDI return levels. 5, 50 and 95 percentiles are shown. (Lower panel:) The change in return levels along with estimated uncertainties based on error propagation. This Figure should be compared to Figure~\ref{fig:rls_1_and_101} to see the effect of including non-stationarity in the GEV-modelling of extremes. The asterisk highlights the return levels at return period 100 years.}
    \label{fig:rl_GEV}
\end{figure}

Figure~\ref{fig:rl_GEV} shows the return levels estimated for years AD 2000 to AD 2100 in the scenarios, along with the KDI return levels (taken from~\cite{kdi2017}). The changes in return levels centre around 200 cm, with considerable upper and lower uncertainty limits. The contributions to the change are about 80cm from the increased msl, 20 cm from increased tide amplitude and 100 cm due to the growth in the location-parameter in the GEV model for extremes.

\subsubsection{Return level changes based on stochastic sampling}
\begin{figure}[!htp]
\centering

    \includegraphics[width=0.32\textwidth]{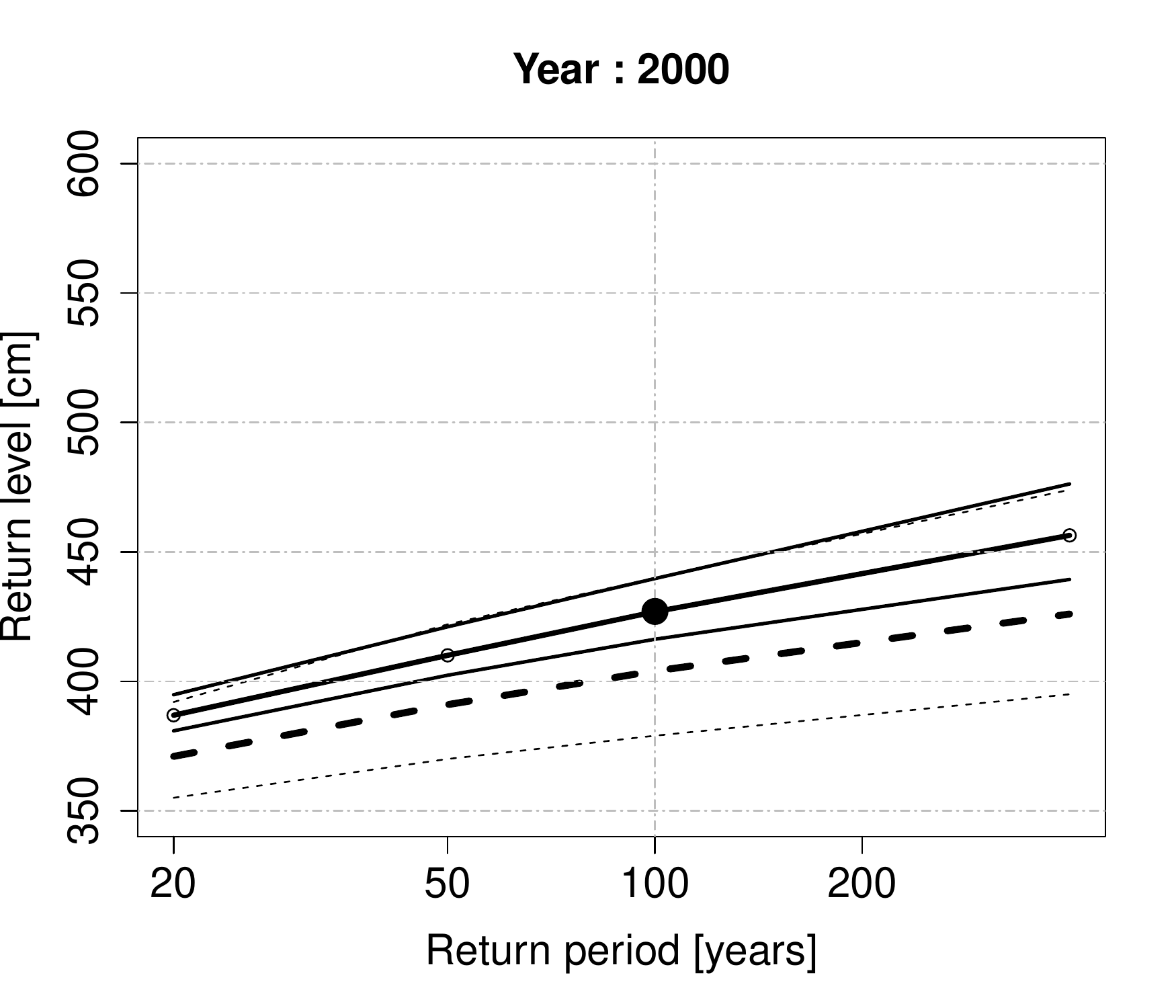}
    \includegraphics[width=0.32\textwidth]{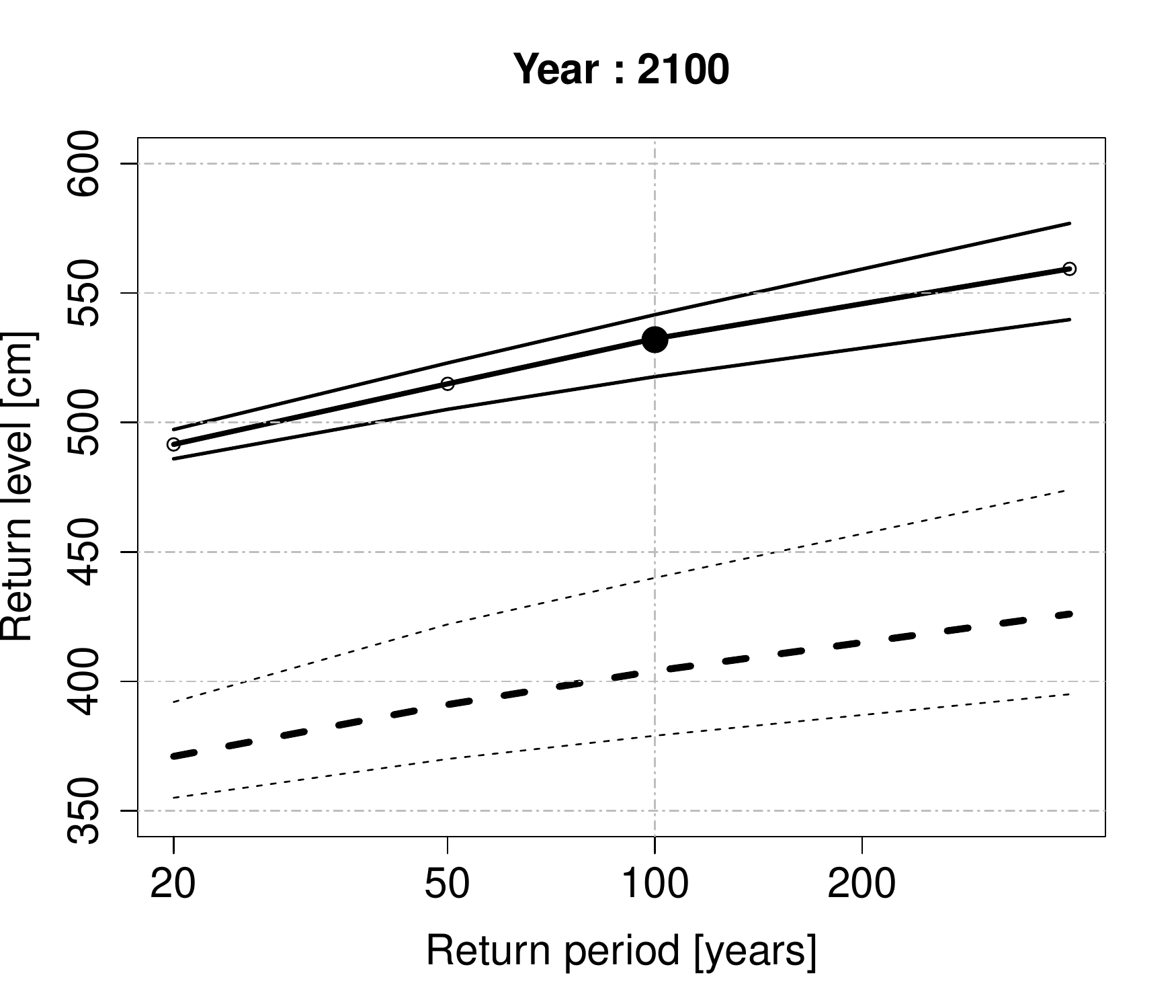}
    \includegraphics[width=0.32\textwidth]{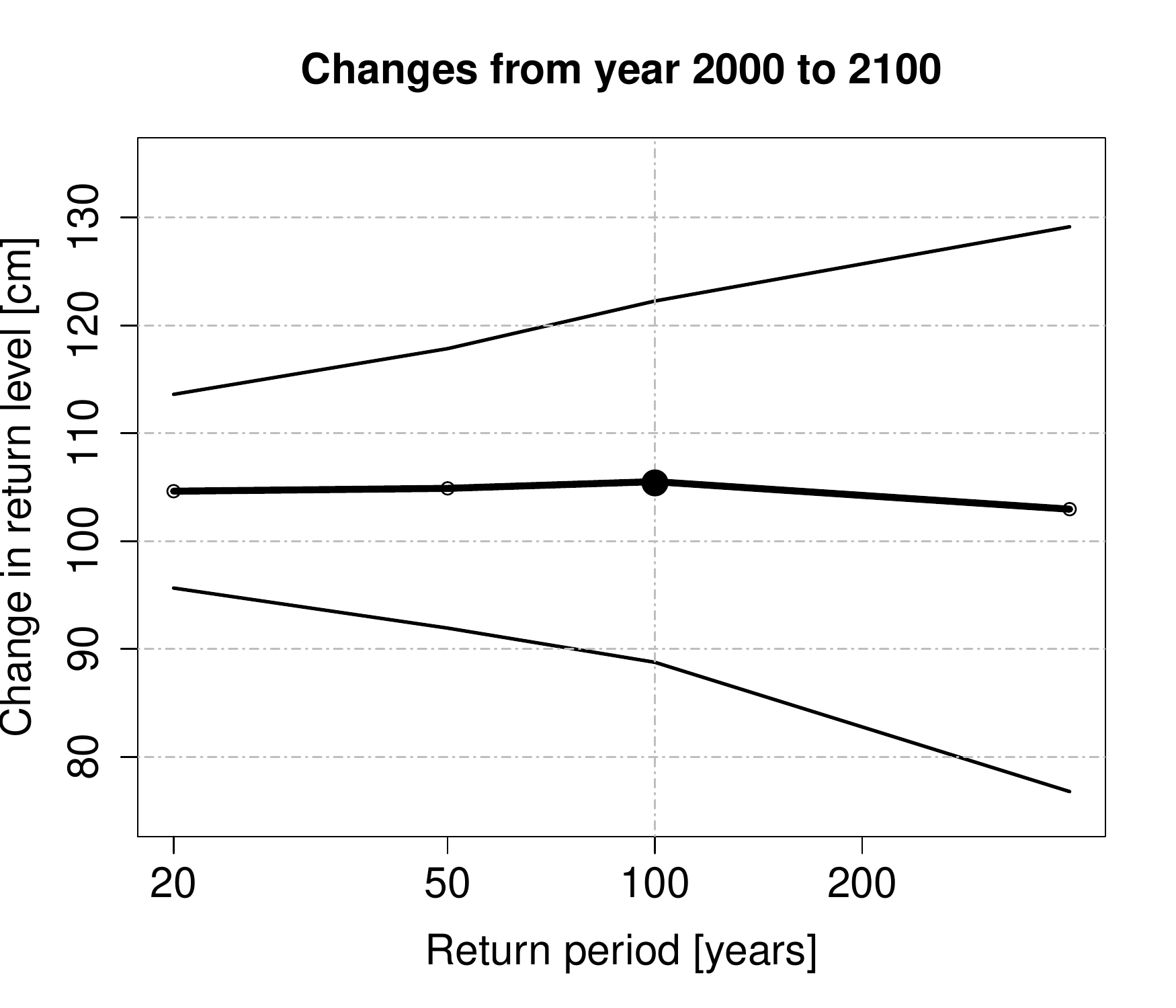}
    \caption{Return levels for the start and the end of the 101-year projected future, based on stochastic sampling of the empirical (stationary) distribution of residuals. This Figure helps understand the effect of the non-stationary distribution of extremes which is used in Figure~\ref{fig:rls_1_and_101}. (\textbf{Left panel:}) Return levels for year 1 of the simulated future. With dashed lines are shown the 5, 50, and 95 percentiles of the log-Normal based estimates of return levels at Esbjerg, from Kystdirektoratet. With solid lines are our own 5, 50 and 95 percentile estimates of near-future return levels. (\textbf{Middle panel:}) Return levels for year 2100 of the scenario period, along with present-day KDI percentiles. (\textbf{Right panel:}) Change in expected return levels, from year 2000 (similar to present) to year 2100, 5th and 95th percentiles shown based on error propagation assuming the uncertainties for year 2000 and year 2100 are independent. The heavy dot at return period 100 years indicates the median return level.}
    \label{fig:rls_1_and_101}
\end{figure}

Figure~\ref{fig:rls_1_and_101} shows return levels for the years 2000 and 2100 of the scenario period using the approach of randomly sampling the 20th century hourly residuals and adding them to $msl$ and the tidal model. For year 1 we match the KDI estimates of return levels, inside the error limits given by KDI. For year 2100 the projected return levels rise considerably. The difference in return levels for year 2000 and 2100 of the projected period is near 105 cm, with the upper limit on the 95 percentile uncertainty level at 130 cm.

\section{Discussion}

We have detected a rise in tide amplitude for some tide constituents, at Esbjerg, and in a purely statistical sense extrapolated the sea level rise vs tide-amplitude relationship. Is this a realistic assumption?
Changes in shelf tide amplitudes have been estimated by others, e.g.~\cite{pickering2012impact} who estimates, on the basis of numerical modelling, a 10 cm rise of the M2 amplitude at Ribe, near Esbjerg (Figure~\ref{fig:mapofDenmark}), and 29 cm at Cuxhaven, for a 2m mean sea level rise. This is an amplitude response rate of 0.05 to 0.15 cm/cm. Our purely regression-based rate estimate is at 0.23$\pm0.04$ cm/cm for M2 amplitude at Esbjerg. \cite{Arns2015} found, also on the basis of numerical modelling, a tide-amplitude response rate of 0.19 cm/cm for the German Bight. Contrary to these findings is~\cite{idier2017sea} who finds a negative response along the Danish West coast for sea level rise less than +5m. Whether the numerical modelling allows for coastal flooding or not is important for the outcome. It would thus seem that recent numerical modelling studies of shelf tides in the Wadden Sea can disagree, and our assumption of continued growth in tide amplitudes with respect to sea level is at the very least not unanimously echoed in the literature -- the possibility of this happening, however, is at the very core of what risk assessment has to contend with. 

Tide amplitudes rising along with a rising sea would contribute to the flood risk at coasts, such as at Esbjerg, and we present our results as being useful for worst-case risk assessments. Improvements in the complex field of realistic numerical modelling of shelf tides could thus have an important impact on this field along the Danish West coast.

A very rough estimate of the expected rise in return levels can be based on the expected rise in sea-levels. From the $msl$ simulations of the future we calculate a mean difference in mean sea level between year 2000 and 2100 of 78 cm (5-50-95 percentiles are 59, 78 and 97 cm respectively). This $msl$ rise should induce a rise in tide amplitude of roughly one fifth of the $msl$ rise (see Table~\ref{tab:constituent_trends_msl}; the line for M2), or 18 cm. This suggests that mean sea levels will rise by 97 cm in 100 years, with a comparable rise in return levels, if we ignore changes in the surges.

Our stochastic simulation based on sampling of 20th century hourly residuals indicates a median rise in return levels, for return periods 20, 50, 100 and 400 years to be near 105 cm, in good agreement with the above simplistic estimate. 

Our other approach, based on non-stationary GEV-modelling of annual maxima formed around msl-dependent tides and a $msl$-dependent location parameter in the GEV distribution, indicates higher rises--the median return levels could rise by 200 cm to some 655 cm, and has a large spread inside the 5-95 percentile band. Towards the end of the 21st century the change in the upper uncertainty limit on return levels reaches 350 cm.

This powerfully illustrates the need for non-stationary modelling of not only future mean sea-levels but also the combined tides and storm surges.

We estimate return level rises while the~\cite{Grinsted2015} and~\cite{Bamber2019}  papers estimate rises in mean sea levels. In the simplest approach a rise in mean sea level implies the same rise in return levels, but with non-stationarity in tides and extreme value distributions such as we have illustrated here, the enhanced rise in tide range and extremes becomes accessible.

Historically, such extreme water levels as we discuss here are not unknown--in terms of actually numerically measured high levels there have been a number of events at Esbjerg, or in the general Danish area of the Wadden Sea, that have gotten close: in AD 1634 water was recorded at +6.3m, in 1981 it rose to +4.3 m, and in 1999, at low tide, the water rose to 4 m at Esbjerg and 5.1 m at nearby Ribe, or just 30 cm below the top of the sea wall. Had high tide coincided with this storm surge then levels at Esbjerg could have risen to 5.4 - 5.5 m, which is far above the current infrastructure  design level at Esbjerg Harbour.

In Norway, the expected future rate of extreme surges in the sea level has been evaluated at various locations along the coastline~\citep[Table 7.5]{NorskRapport2015}. At Vestlandet, the "return level corresponding to the present 200-year event will be exceeded about 40 times during the 21st century". We can calculate a similar statistic from our non-stationary simulation of Esbjerg levels. We find that levels will exceed the level corresponding to the present 200-year event 8-15 times during the next 100 years, with most of the events happening late in the century--e.g. half of the events happen in the last 18 to 27 years of the 21st century, in the simulations. This is less frequently than at fjord-based sites in Vestland, but it is more than at Heimsjø and Tromsø, which are further up north on the west coast.

The North Atlantic Oscillation index has a degree of control over where low pressure systems cross the North Atlantic and make landfall -- suggesting that the NAO index may have an influence over local changes in storminess along the Danish western coastline. At an early stage of the present work we included the NAO index as a co-variate and found a suggestion that it was a co-variate to $GEV$ scale, but set the suggestion aside for future analysis as the statistical significance was marginal.

\section{Conclusions}
Our projected median rise in return levels at Esbjerg is near 200 cm, for return periods from 20 to 200 years. Non-stationary methods predict a 100 cm larger rise in return levels than does the extremes-stationary method, for all considered return periods. Additionally, the upper uncertainty limits increase when taking non-stationarity into account. This highlights the importance of using non-stationary methods for projections of worst-case events.

%\section{Perspectives}
%\include{perspective
\textbf{Acknowledgements}\\
We acknowledge use of hourly GESLA data from the danish station at Esbjerg; https://www.gesla.org/, and the use of data~\citep{PSMSL2012} from the
Permanent Service for Mean Sea Level (PSMSL), 2019, "Tide Gauge Data", Retrieved 30 May 2019 from http://www.psmsl.org/data/obtaining/. Suggestions and comments from Jian Su, at DMI, are warmly acknowledged.

%\end{acknowledgements}

%% Figures

\bibliographystyle{spbasic}      % basic style, author-year citations
%\bibliography{references}   % name your BibTeX data base


\end{document}